\documentclass[conference]{IEEEtran}
\IEEEoverridecommandlockouts
\usepackage{cite}
\usepackage{amsmath,amssymb,amsfonts}
\usepackage{amsthm}
\usepackage{bm}
\usepackage{mathtools}
\usepackage{algorithm}
\usepackage{algorithmic}
\usepackage{graphicx}
\usepackage{textcomp}
\usepackage{xcolor}
\usepackage{dsfont}
\usepackage{verbatim}
\newtheorem{theorem}{Theorem}
\newtheorem{theorem*}{Theorem*}

\newtheorem{example*}[theorem*]{Example}
\newtheorem{definition}[theorem]{Definition}
\newtheorem{corollary}[theorem]{Corollary}
\newtheorem{proposition}[theorem]{Proposition}

\newcommand{\bmf}{\bm{f}}

\newcommand{\bv}{\bm{v}}    

\newcommand{\btheta}{\bm{\theta}}
\newcommand{\bTheta}{\bm{\Theta}}    
\newcommand{\bmu}{\bm{\mu}}    

\newcommand{\bphi}{\bm{\phi}}

\newcommand{\cX}{\mathcal{X}}

\newcommand{\defeq}{\triangleq} 
\newcommand{\reals}{\mathbb{R}}
\newcommand{\T}{\mathrm{T}}    

\DeclareMathOperator*{\argmax}{arg\,max}
\DeclareMathOperator*{\argmin}{arg\,min}

\DeclareMathOperator{\tr}{tr}

\DeclareMathOperator{\E}{\mathbb{E}}
\DeclareMathOperator{\Prob}{\mathbb{P}}

\newcommand{\secref}[1]{Section~\ref{#1}}

\newcommand{\figref}[1]{Figure~\ref{#1}}

\newcommand{\thmref}[1]{Theorem~\ref{#1}}
\newcommand{\propref}[1]{Proposition~\ref{#1}}

\newcommand{\algoref}[1]{Algorithm~\ref{#1}}

\def\BibTeX{{\rm B\kern-.05em{\sc i\kern-.025em b}\kern-.08em
    T\kern-.1667em\lower.7ex\hbox{E}\kern-.125emX}}
\begin{document}

\title{An Information-theoretic Method for Collaborative Distributed Learning with Limited Communication
	
}

\author{
	{\rm Xinyi Tong, Jian Xu, Shao-Lun Huang} \\
	Tsinghua-Berkeley Shenzhen Institute, Tsinghua University\\
	\{txy18, xujian20\}@mails.tsinghua.edu.cn, shaolun.huang@sz.tsinghua.edu.cn
}


\maketitle

\begin{abstract}

In this paper, we study the information transmission problem under the distributed learning framework, where each worker node is merely permitted to transmit a $m$-dimensional statistic to improve learning results of the target node. Specifically, we evaluate the corresponding expected population risk (EPR) under the regime of large sample sizes. We prove that the performance can be enhanced since the transmitted statistics contribute to estimating the underlying distribution under the mean square error measured by the EPR norm matrix. Accordingly, the transmitted statistics correspond to the eigenvectors of this matrix, and the desired transmission allocates these eigenvectors among the statistics such that the EPR is minimal. Moreover, 
we provide the analytical solution of the desired statistics for single-node and two-node transmission, where a geometrical interpretation is given to explain the eigenvector selection. For the general case, an efficient algorithm that can output the allocation solution is developed based on the node partitions.

\end{abstract}


\section{Introduction}
Distributed learning has been an active research area focusing on solving learning tasks of different workers under the collaboration between each other \cite{VerbraekenWKKVR20, li2020distributed, LiuHZLJXD22}. This learning scheme allows the distributed workers to share some knowledge that enables them to collaboratively learn better models than learning individually. In this context, the communication cost is of significant importance as the worker nodes usually have power and bandwidth resource constraints in realistic applications \cite{ImteajTWLA22}. On the other hand, the learning tasks are usually solved by iterative optimization steps, e.g., stochastic gradient decent (SGD) \cite{Bottou10}, which could involve iterative high-dimensional gradient message transmissions in a frequent manner and thus induce high communication overhead. 

To alleviate this issue, recent works have studied the approaches of gradient compression, which focus on employing a low-precision and low-dimensional representation of the gradient vector \cite{h17qsgd,Wangni18sparse,JiangA18,BasuDKD20}. On the other hand, some other works employ multiple local learning steps before transmitting gradient to reduce the total communication rounds \cite{Stich19,SpiridonoffOP21,GorbunovHR21}. However, those methods should empirically strike a good balance between the model performance and the communication cost, and they did not consider the explicit constraint of communication bits. Therefore, designing the transmission approach with limited communication, i.e., revealing what statistics are required to be transmitted from other nodes to the target one, is vital for taking full advantage of the distributed collaborative learning~\cite{dis}.

In this paper, we study the fundamental problem of information transmission in distributed learning under the information dimensionality constraint, where the setting is summarized as follows. Let $X$ be the random variable of data with domain $\cX$. We consider that the distributed learning problem has $k+1$ worker nodes, namely node $0$, node $1$, \dots , node $k$. For node $i$, we assume that $n_i$ training samples $\{x^{(i)}_{j}\}_{j=1}^{n_i}$ are i.i.d. generated from the universal distribution $P_X$, where $\hat{P}_X^{(i)}$ denotes the corresponding empirical distribution. In detail, the training process follows the empirical risk minimization (ERM) framework, where each node learns the parameter vector $\btheta\in \reals^{D}$ with respect to the loss function $l(x;\btheta)\in \reals$.

\begin{figure}[t]
	\vskip -0.25in
	\centerline{\includegraphics[width=\linewidth]{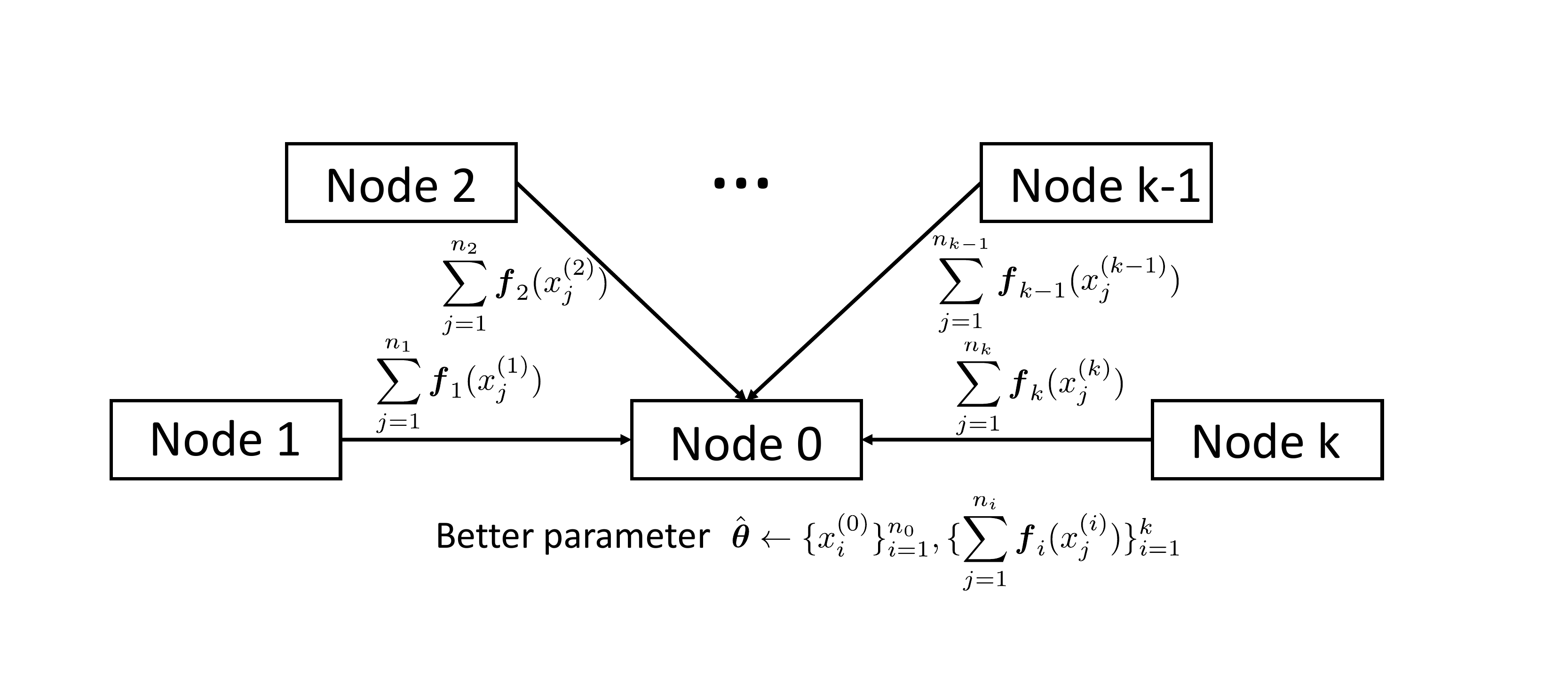}}	
		\vskip -0.25in
	\caption{A geometric explanation of distributed learning setting, where each node transmit a statistic to the target node $0$ to help achieve a better parameter.}
	\label{fig:0}
		\vskip -0.1in
\end{figure}  

Without loss of generality, we take node $0$ as the target. As shown in \figref{fig:0}, the distributed learning framework can achieve a better parameter vector for node $0$ by the following collaboration mechanism. For all the remaining nodes, node $i$ ($i=1,2,\cdots,k$) transmits a $m$-dimensional statistic to node $0$, which takes the empirical mean of some statistic function $\bmf_i\colon \cX\rightarrow \reals^m$ of the samples, i.e., its form is $\sum_{j=1}^{n_i}\bmf_i(x^{(i)}_j)=\sum_{x\in\cX}\hat{P}_X^{(i)}(x)\bmf_i(x)$. The restriction on the dimensionality is typically a communication constraint of fixed codeword length when each dimension needs fixed transmission bits.

The purpose of our work is to provide the expressions of the statistic functions $\{\bmf_i(\cdot)\}_{i=1}^k$ such that the learning result $\hat{\btheta}$ of node $0$ performs best, where $\hat{\btheta}$ can be recognized as a function of its individual samples and $\{\sum_{j=1}^{n_i}\bmf_i(x^{(i)}_j)\}_{i=1}^k$. Moreover, the performance is evaluated by the expected population risk (EPR), i.e., the desired statistic functions can be derived by
\begin{align}
	\label{eq:lok}
	\min_{\{\bmf_i(\cdot)\}_{i=1}^k}\ \E\left[\sum_{x\in\cX}P_X(x)l(x;\hat{\btheta})\right],
\end{align}
where the expectation is taken over the sampling process. In this paper, we consider the asymptotic regime that the sample size of each node is large, and the empirical distribution can be close to the underlying distribution with high probability. As a result, we demonstrate that the EPR can be recognized as a mean square error measured by the EPR norm matrix between the empirical and underlying distribution.   

Note that the empirical distribution can be regarded as a Gaussian vector under the asymptotic regime, whose covariance is inversely proportional to the sample size. Accordingly, we prove that the statistic functions $\{\bmf_i(\cdot)\}_{i=1}^k$ correspond to the eigenvectors of the EPR norm matrix. Therefore, designing the optimal information transmission mechanism is transformed into an integer programming problem, which settles different eigenvectors to the positions of different statistic functions. Especially, we provide the analytical solutions of the cases when $k=1$ and $k=2$, where eigenvectors of the largest 2 eigenvalues are allocated and a geometric interpretation is given. Moreover, we demonstrate that the statistic functions of those nodes with more samples prefer the eigenvectors with larger eigenvalues. This conclusion leads to an algorithm based on the partition of the $k$ nodes, which presents a computational complexity smaller than trivial methods.

Our framework and results differ from previous works in two ways: (1) previous works transmit the compressed gradient vectors in each round, while we transmit the low-dimensional statistics; (2) previous works involve iterative gradient transmissions, while our goal is to maximize the utility of collaboration between workers by one-shot communication. The contributions of this paper can be summarized as follows. \secref{sec:2} formulates this problem as estimating the underlying distribution under EPR. \secref{sec:3} presents the main theorems describing the properties of the solutions for scalar transmission, and propose the algorithm based on node partitions. Finally, \secref{sec:4} extends the results to the case of vector transmission and improves the algorithm according to the operation of high order partitions.
\section{Preliminaries}
\label{sec:2}

\subsection{Asymptotic Approximation}
Before presenting the problem formulation, we briefly introduce some convergence results and notations with respect to the empirical distributions. First, for an arbitrary distribution $Q_X$, we define its associated information vector $\bphi$ as, 
\begin{align}
	\bphi(x)=\frac{Q_X(x)}{\sqrt{P_X(x)}},\ \forall x\in \cX,
\end{align}
denoted as $\phi\leftrightarrow Q_X$. Accordingly, information vector $\bphi^*\leftrightarrow P_X$ follows $\bphi^*(x)=\sqrt{P_X(x)}$, and $\hat{\bphi}_i\leftrightarrow \hat{P}^{(i)}_X$ is associated with the corresponding empirical distribution.


In this paper, we concentrate on the local regime where 
\begin{align}
	\label{eq:ap}
	Q_X(x)-P_X(x)=O(\epsilon),
\end{align}
for all $x$ and $\epsilon$ is small. The empirical distributions are contained in this regime with high probability when samples sizes are large. Under this regime, we have the following approximation of the Kullback-Leibler (K-L) divergence 
\begin{align}
	\label{eq:lc}
	D(Q_X\|P_X)=\frac{1}{2}\left\|\bphi-\bphi^*\right\|^2+O(\epsilon^2),
\end{align}
where $\|\cdot\|^2$ denotes the $l_2$-norm of vectors. 

It is well known in information theory \cite{csi} that the probability function follows
\begin{align*}
	\mathbb{P}\left(\hat{\bphi}_i;\bphi^*\right)=\mathbb{P}\left(\hat{P}_X^{(i)};P_X\right)\doteq \exp\left(-n_iD(\hat{P}_X^{(i)}\|P_X)\right).
\end{align*}

By applying the local approximation \eqref{eq:lc}, the probability function follows
\begin{align}
	\mathbb{P}\left(\hat{\bphi}_i;\bphi^*\right)\doteq \exp\left(-\frac{n_i}{2}\left\|\hat{\bphi}_i-\bphi^*\right\|^2+O(\epsilon^2)\right),
\end{align} 
 which indicates that $\hat{\bphi}_i$ is approximately a Gaussian vector centered at $\bphi^*$ with covariance matrix $\frac 1{n_i} \mathbf{I}$.
 
 \subsection{Computation of EPR}
 Note that the optimal parameter that all the nodes desire can be defined as
 \begin{align}
 	\label{eq:ffdd}
 	\btheta^*\defeq \argmin_{\btheta\in\reals^{D}}\sum_{x\in\cX}P_X(x)l(x;\btheta).
 \end{align}
  Consider that we learn the estimator $\hat{\btheta}$ for $\btheta^*$ with respect to $Q_X$, which is typically the empirical distribution of the corresponding node when no knowledge is transferred from other nodes, i.e., let the learned estimator be
\begin{align}
\label{eq:ffff}
	\hat{\btheta}\defeq\argmin_{\btheta\in\reals^{D}}\sum_{x\in\cX}Q_X(x)l(x;\btheta).
\end{align}
The performance of this estimator is evaluated by the expected population risk (EPR), which is defined as [cf. \eqref{eq:lok}]
\begin{align}
	\label{eq:kj}
	R(\hat{\btheta})\defeq \E_{\mathtt{G}}\left[\sum_{x\in\cX}P_X(x)l(x;\hat{\btheta})\right]-\sum_{x\in\cX}P_X(x)l(x;\btheta^*),
\end{align}
where the expectation is computed by the integral over the Gaussian density functions
\begin{align}
	\label{eq:gau}
		\mathbb{P}\left(\hat{\bphi}_i;\bphi^*\right)= \left(\frac{2\pi}{n_i}\right)^{-\frac{|\cX|}{2}}\cdot \exp\left(-\frac{n_i}{2}\left\|\hat{\bphi}_i-\bphi^*\right\|^2\right),
\end{align}
, and $\sum_{x\in\cX}P_X(x)l(x;\btheta^*)$ (constant) is the EPR where the optimal parameter $\btheta^*$ is achieved. Based on this formulation, we have the following characterization of the EPR \eqref{eq:kj}.

\begin{proposition}\label{lem:2}
	Suppose that $l(x;\btheta)$ is twice-differentiable and Lipschitz continuous for $\btheta$, and $\hat{\btheta}$ is unbiased for $\btheta^*$, the testing loss as defined in \eqref{eq:kj} can be computed by
	\begin{align}
	\label{eq:ujh}
			R(\hat{\btheta})=\frac{1}{2}\mathbb{E}_{\mathtt{G}}\left[\left(\bphi-\bphi^*\right)^T\mathbf{H}\left(\bphi-\bphi^*\right)\right]+o\left(\epsilon^2\right),
	\end{align}
	where $\mathbf{H}\in \reals^{|\cX|\times|\cX|}$ is called \textbf{EPR norm matrix}, whose entries are
	\begin{align*}
		\mathbf{H}(x_1,x_2)=\sqrt{P_X(x_1)P_X(x_2)}\nabla_{\btheta}l(x_1;\btheta^*)^\T\bm{\Theta}^\dagger\nabla_{\btheta}l(x_2;\btheta^*)
	\end{align*}
and 
\begin{align}
	\bm{\Theta}=\sum_{x\in\cX}P_X(x)\nabla^2_{\btheta}l(x;\btheta^*).
\end{align}
The notation $\nabla_{\btheta}l(x_2;\btheta^*)$ and $\nabla^2_{\btheta}l(x_2;\btheta^*)$ denote the gradient vector $\left.\nabla_{\btheta}l(x_2;\btheta)\right|_{\btheta=\btheta^*}$ and Hessian matrix $\left.\nabla^2_{\btheta}l(x_2;\btheta)\right|_{\btheta=\btheta^*}$, and $\bTheta^\dagger$ denotes its Moore–Penrose inverse.
\end{proposition}
This characterization indicates that the purpose of our problem is to find the optimal estimation for $\bphi^*$ with its Gaussian observations under error \eqref{eq:ujh}, which can be seen as a mean square error measured by $\mathbf{H}$.

 When no knowledge is transferred from other nodes, node $0$ takes $\bphi=\hat{\bphi}_0$, which has the EPR (high order terms omitted)
\begin{align}
	\label{eq:poi}
	\frac{1}{2}\mathbb{E}_{\mathtt{G}}\left[\left(\hat{\bphi}_0-\bphi^*\right)^T\mathbf{H}\left(\hat{\bphi}_0-\bphi^*\right)\right]=\frac{\tr(\mathbf{H})}{2n_0}.
\end{align}

When $\{\sum_{x\in\cX}\hat{P}_X^{(i)}(x)\bmf_i(x)\}_{i=1}^k$ are transmitted from other nodes, this paper could construct a smaller EPR than \eqref{eq:poi}. Let
\begin{align*} 
\bm{F}_i\defeq \left[\sqrt{P_X(1)}\bmf_i(1),\cdots,\sqrt{P_X(|\cX|)}\bmf_i(|\cX|)\right]\in \reals^{m\times|\cX|}
\end{align*}
 be the statistic function matrix and the statistic from node $i$ can be written as $\bm{F}_i\hat{\bphi}_i$. Finally, the problem \eqref{eq:lok} comes into an optimization problem with two steps: 
\begin{enumerate}
	\item[(i)] provide the optimal estimator $\tilde{\bphi}$ with respect to the empirical vector $\hat{\bphi}_0$ and the statistics $\{\bm{F}_i\hat{\bphi}_i\}_{i=1}^k$;
	\item[(ii)] find the optimal $\bm{F}_i$'s such that the EPR is minimal.
\end{enumerate}
Thus, the following formulation is given

\begin{align}
	\label{eq:que}
	\min_{\{\bm{F}_i\}_{i=1}^k}\  \min_{\tilde{\bphi}(\hat{\bphi}_0,\{\bm{F}_i\hat{\bphi}_i\}_{i=1}^k)}\mathbb{E}_{\mathtt{G}}\left[(\tilde{\bphi}-\bphi^*)^T\mathbf{H}(\tilde{\bphi}-\bphi^*)\right].
\end{align}
\section{Scalar Transmission}
\label{sec:3}
In this section, we provide the solution of problem \eqref{eq:que} under a special case when each node only transmits a scalar to node $0$, i.e., $m=1$. In other word, the matrix $\bm{F}_i$ is degenerated to a vector, which is denoted as $\bmu_i^\T$. This special case can be easily extended to the case when $m>1$, and the result would be shown in \secref{sec:4}.

First, we provide the solution for step (i). Let $\tilde{\bphi}^*$ be the optimal estimator that minimizes the EPR
\begin{align}
	\label{eq:pol}
	\tilde{\bphi}^*=\argmin_{\tilde{\bphi}(\hat{\bphi}_0,\{\bmu_i^\T\hat{\bphi}_i\}_{i=1}^k)}\mathbb{E}_{\mathtt{G}}\left[(\tilde{\bphi}-\bphi^*)^\T\mathbf{H}(\tilde{\bphi}-\bphi^*)\right],
\end{align}
which is almost a non-Bayesian minimal mean square error (MMSE) estimation problem. Note that $(\tilde{\bphi}-\bphi^*)^\T\mathbf{H}(\tilde{\bphi}-\bphi^*)=\|\mathbf{H}^{\frac 12}(\tilde{\bphi}-\bphi^*)\|^2$, where $\mathbf{H}^{\frac 12}$ satisfies $\mathbf{H}^{\frac 12\T}\mathbf{H}^{\frac 12}=\mathbf{H}$. Thus, problem \eqref{eq:pol} can be viewed as to find the MMSE estimator for the linearly transformed parameter $\mathbf{H}^{\frac 12}\bphi^*$. Note that it is easy to verify that the corresponding observations $\mathbf{H}^{\frac 12}\hat{\bphi}_i$'s are still Gaussian vectors. The typical method is to compute the maximum-likelihood estimator (MLE) and then prove its efficiency by the Cramer-Rao bound. The MLE $\tilde{\bphi}_{\mathtt{ML}}$ can be computed as follows
\begin{align}
		\mathbf{H}^{\frac 12}\tilde{\bphi}_{\mathtt{ML}}=\argmax_{\mathbf{H}^{\frac 12}\bphi^*}\ \Prob\left(\hat{\bphi}_0;\bphi^*\right)\prod_{i=1}^k\Prob\left(\bmu_i^\T\hat{\bphi}_i;\bphi^*\right),
	\end{align}
where the density function $\Prob\left(\hat{\bphi}_i;\bphi^*\right)$ is defined in \eqref{eq:gau}.

Accordingly, the expression of $\tilde{\bphi}_{\mathtt{ML}}$ is 
\begin{align}
	\label{eq:tre}
	\tilde{\bphi}_{\mathtt{ML}}=\left(n_0\mathbf{I}+\sum_{i=1}^k n_i\frac{\bmu_i\bmu_i^\T}{\bmu_i^\T\bmu_i}\right)^{-1} \left(n_0\hat{\bphi}_0+\sum_{i=1}^k\frac{\bmu_i\bmu_i^\T\hat{\bphi}_i}{\bmu_i^\T\bmu_i}\right).
\end{align}
Then, we have the following characterization of the optimal estimator $\tilde{\bphi}^*$.
\begin{theorem}
	\label{thm:1}
	The optimal estimator
	$\tilde{\bphi}^*$ as defined in \eqref{eq:pol} takes the form of the MLE $\tilde{\bphi}_{\mathtt{ML}}$ as defined in \eqref{eq:tre}.
\end{theorem}
The next step is to compute the corresponding EPR $\mathbb{E}_{\mathtt{G}}[(\tilde{\bphi}_{\mathtt{ML}}-\bphi^*)^T\mathbf{H}(\tilde{\bphi}_{\mathtt{ML}}-\bphi^*)]$.
Without loss of generality, we assume that the statistic functions satisfy $\bmu_i^\T\bmu_i=1$ ($i=1,2,\cdots,k$), and the step (ii) of problem \eqref{eq:que} becomes 
\begin{align}
	\label{eq:nhg}
	\min_{\{\bmu_i\}_{i=1}^k}&\ \tr\left[\mathbf{H}\left(n_0\mathbf{I}+\sum_{i=1}^kn_i\bmu_i\bmu_i^\T\right)^{-1}\right]\\ \notag
	\mathrm{s.t.}&\ \bmu_i^\T\bmu_i=1.
\end{align}
The following theorem characterizes the property of the solution of this problem.
\begin{theorem}
	\label{thm:2}
	Suppose that the eigenvalues and the corresponding eigenvectors of matrix $\mathbf{H}$ as defined in \propref{lem:2} are $\{\lambda_i\}_{i=1}^{|\cX|}$ and $\{\bv_i\}_{i=1}^{|\cX|}$, where $\lambda_1\ge \lambda_2\ge \cdots\ge \lambda_{|\cX|}\ge 0$. Let $\{\bmu_i^*\}_{i=1}^k$ be the optimal arguments of \eqref{eq:nhg}, and then
	\begin{align}
		\bmu_i^*\in \{\bv_1,\bv_2, \cdots,\bv_{|\cX|}\},\ \forall i\in\{1,2,\cdots,k\}.
	\end{align}
\end{theorem}
\thmref{thm:2} indicates that the statistic design searches for  the suitable eigenvectors such that the EPR is minimal. Let  $\{c_i\}_{i=1}^{k}$ be the index set such that $\bmu_i=\bv_{c_i}$, and then problem \eqref{eq:nhg} becomes
\begin{align}
	\label{eq:oiu}
	\min_{c_i\in\{1,2,\cdots,|\cX|\}}\ \sum_{j=1}^{|\cX|} \frac{\lambda_j}{n_0+\sum_{i=1}^{k}\mathds{1}\{c_i=j\}n_i},
\end{align}
where $\mathds{1}\{\cdot\}$ denotes the indicator function \cite{indicator}.

Note that problem \eqref{eq:oiu} is an integer programming problem, which is typically NP-hard~\cite{nphard}, and the analytical solution is hard to provide. However, we can still understand some properties of its solution and provide an efficient algorithm. Before presenting these results, we first show two simple cases of this problem, which could help achieve a geometrical understanding and interpretation. Specifically, when $k=0$, the objective function of problem \eqref{eq:oiu} gives $\sum_{j=1}^{|\cX|}\lambda_j/n_0$, which is consistent with the result in \eqref{eq:poi}.

\begin{figure}[t]
	\centerline{\includegraphics[width=\linewidth]{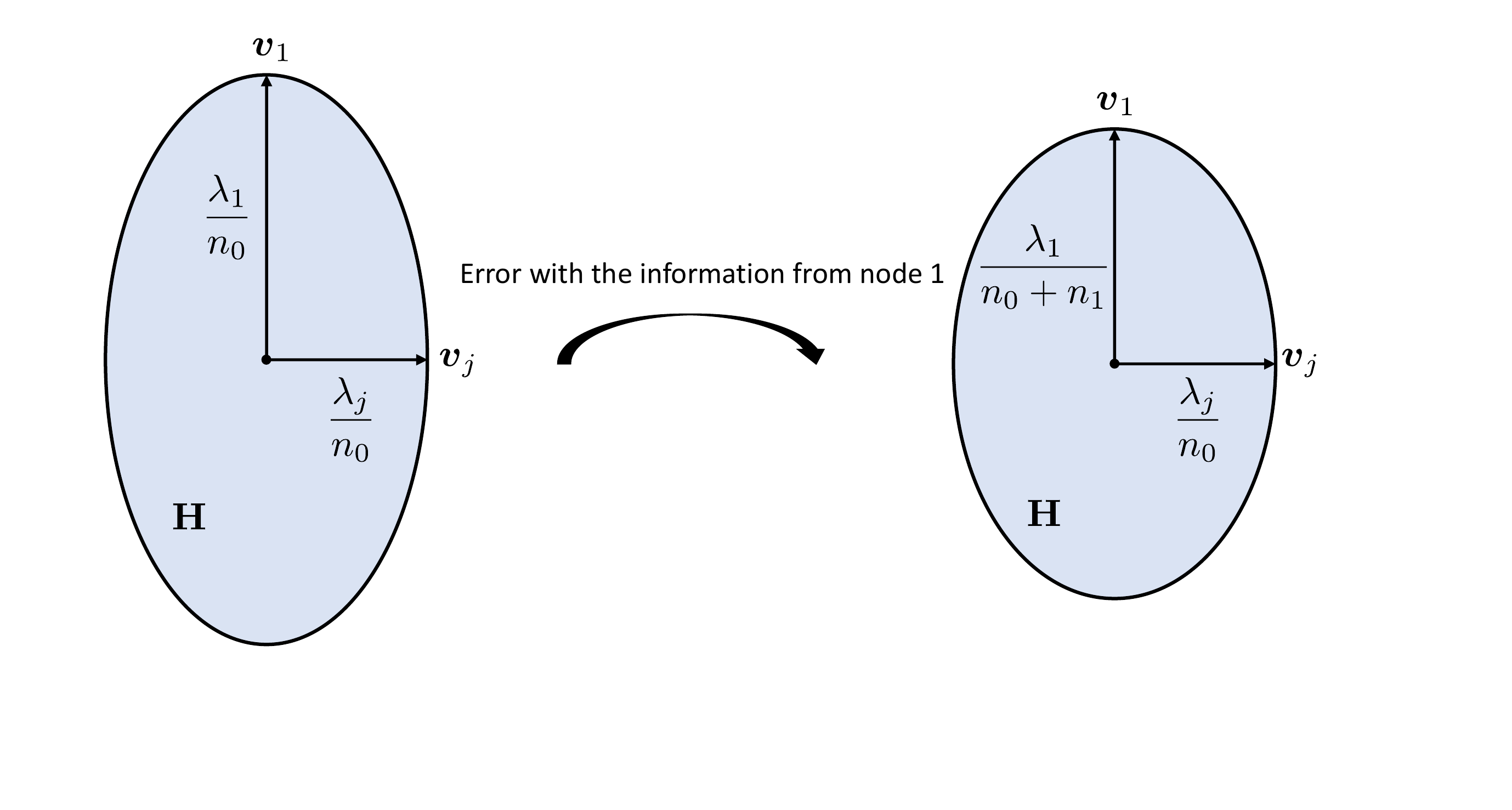}}
		\vskip -0.25in	
	\caption{A geometric illustration of Example 1, where the blue shadow implies the EPR along different directions in the space of information vectors. The information transmission deduces the expected error along $\bv_1$ from $\lambda_1/n_0$ to $\lambda_1/n_0+n_1$.}
	\label{fig:1}
		\vskip -0.1in
\end{figure}

\begin{figure}[t]
	\centerline{\includegraphics[width=0.9\linewidth]{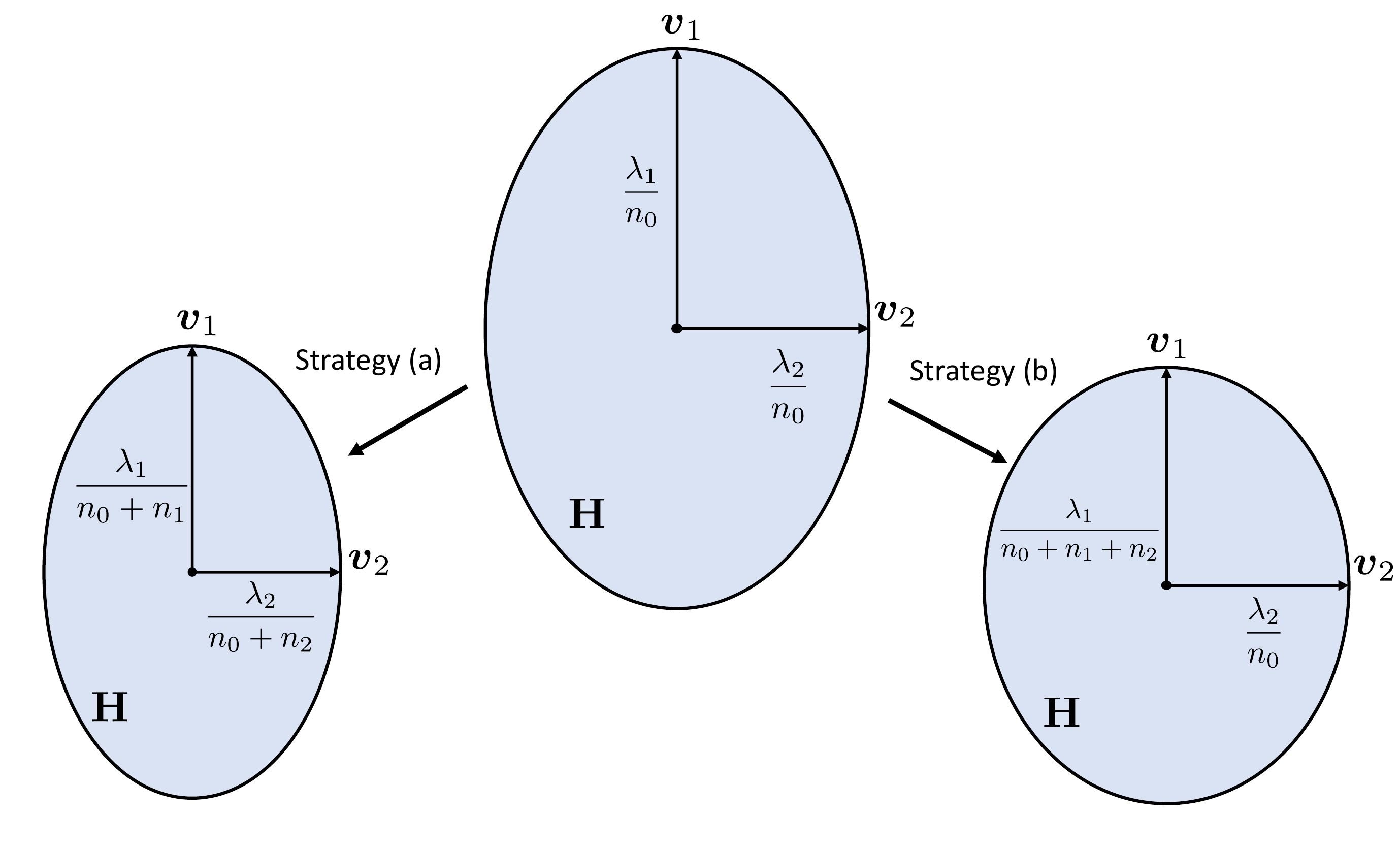}}
	\vskip -0.1in
	\caption{A geometric illustration of Example 2, where strategy (a) leads to error deductions in the directions of both $\bv_1$ and $\bv_2$, while strategy (b) in the direction of merely $\bv_1$.}
	\label{fig:2}
	\vskip -0.1in
\end{figure}

\begin{proposition}[Single-node transmission]
	\label{prop:1}
When there is only one node for knowledge transmission, i.e., $k=1$, problem \eqref{eq:oiu} comes to
\begin{align}
	\label{eq:ggt}
	\min_{c_1\in\{1,2,\cdots,|\cX|\}}  \frac{\lambda_{c_1}}{n_0+n_1}+\sum_{j\ne c_1}\frac{\lambda_j}{n_0}.
\end{align}
Let $c_1^*$ be the solution of problem \eqref{eq:ggt} and it is easy to verify that $c_1^*=1$. Accordingly, the optimal statistic function $\bmu_1^*$ is the largest eigenvector $\bv_1$ of matrix $\mathbf{H}$.

 A geometric explanation associated with this result can be depicted in \figref{fig:1}. Note that the case when $k=0$ implies that the EPR \eqref{eq:poi} is the summation of the expected errors along  all the eigenvectors of matrix $\mathbf{H}$, which are proportional to the corresponding eigenvalues and inversely proportional to the sample size $n_0$. With the information contained in the scalar $\bmu_1^\T \hat{\bphi}_1$, the expected error along $\bmu_1$ is deduced from $\lambda_{c_1^*}/n_0$ to $\lambda_{c_1^*}/n_0+n_1$. Problem \eqref{eq:ggt} aims at finding the direction where the maximum error deduction is achieved, where obviously the direction of $\bv_1$ is the answer.
\end{proposition}
\begin{proposition}[Two-node transmission]
	When there are two nodes for knowledge transmission, i.e., $k=2$, there exist two possible strategies
	\begin{enumerate}
		\item[(a)] Statistic function $\bmu_1$ and $\bmu_2$ select different directions;
		\item[(b)] Statistic function $\bmu_1$ and $\bmu_2$ select the same direction.
	\end{enumerate}
	Then, problem \eqref{eq:oiu} under these two strategies comes to
	\begin{subequations}
	\begin{align}
		&\min_{c_1,c_2\in\{1,2,\cdots,|\cX|\}, c_1\ne c_2} \frac{\lambda_{c_1}}{n_0+n_1}+\frac{\lambda_{c_2}}{n_0+n_2}+ \sum_{j\ne c_1,c_2}\frac{\lambda_j}{n_0}\label{eq:fgt}\\
    	&\min_{c_1\in\{1,2,\cdots,|\cX|\}} \frac{\lambda_{c_1}}{n_0+n_1+n_2}+\sum_{j\ne c_1}\frac{\lambda_j}{n_0}\label{eq:fgtt}		
	\end{align}
\end{subequations}
    Without loss of generality, it could be assumed that $n_1\ge n_2$. The solutions of problem \eqref{eq:fgt} and  \eqref{eq:fgtt} are easy to derive. For strategy (a), the direction of $\bmu_1$ and $\bmu_2$ shall be along $\bv_1$ and $\bv_2$, i.e., the optimal arguments are $c_1^*=1$ and $c_2^*=2$; for strategy (b), similar to \propref{prop:1}, $c_1^*=1$. Additionally, the corresponding EPRs are presented in the following.
\begin{subequations}
	\begin{align}
	&\frac{\lambda_{1}}{n_0+n_1}+\frac{\lambda_{2}}{n_0+n_2} +\sum_{j=3}^{|\cX|}\frac{\lambda_j}{n_0}\label{eq:fgy}\\
	&\frac{\lambda_{1}}{n_0+n_1+n_2}+\sum_{j=2}^{|\cX|}\frac{\lambda_j}{n_0}\label{eq:oibg}
	\end{align}
\end{subequations}
Depending on the relationship between the eigenvalues $\lambda_1$ and $\lambda_2$, the EPR of strategy (a) could be larger or smaller than strategy (b). Thus, the optimal statistic functions of the two nodes are decided by the following test.
\begin{align}
	\frac{\lambda_1}{\lambda_2} \mathop{\gtreqless}^{\mathrm{Strategy\ (b)}}_{\mathrm{Strategy\ (a)}}\limits\frac{(n_0+n_1)(n_0+n_1+n_2)}{n_0(n_2+n_0)}.
\end{align}
A geometric explanation associated with this result can be depicted in \figref{fig:2}. When the largest eigenvalue $\lambda_1$ is sufficiently large, the additional information from $n_1$ samples of node 1 and $n_2$ samples of node 2 tends to reduce the population error along the same direction $\bv_1$, and otherwise the two nodes are allocated to different directions.
\end{proposition}
These two propositions imply that the information transmission corresponds to allocating different directions of eigenvectors to different worker nodes. For a general case $k\ge 2$, the allocation decision depends on the relationship between the eigenvalues of matrix $\mathbf{H}$. As the most trivial way to solve this problem, we could try all possible $c_1,\cdots,c_k$ such that $c_i\in\{1,2,\cdots,k\}$ (not the set $\{1,2,\cdots,|\cX|\}$ since we at most reduce EPR along the direction of $\bv_k$ to achieve a larger EPR deduction), which contains ${k}^{k}$ possible allocations.

However, the complexity can be reduced by considering all possible strategies. As shown in Example 2, when $k=2$, there could be $2^2$ possible allocations but only 2 possible strategies. Moreover, each strategy corresponds to a possible partition of the index set $\{1,\cdots,k\}$. For instance, strategy (a) and (b) in Example 2 corresponds to the partition $\{\{1\},\{2\}\}$ and $\{\{1,2\}\}$. In detail, let $\mathcal{T}=\{t_1,\cdots,t_{|\mathcal{T}|}\}$ be a partition of $\{1,\cdots,k\}$, and the corresponding strategy refers to that the correspondingly indexed statistic functions are the same eigenvector, i.e., for all the elements $a_1, \cdots, a_{|t_i|}\in t_i$, $\bmu_{a_i}=\cdots=\bmu_{a_{|t_i|}}$. Thus, given partition $\mathcal{T}$, problem~\eqref{eq:oiu} becomes
\begin{align}
	\label{eq:asd}
	\min_{\{\lambda^\prime_i\}_{i=1}^{|\cX|}\in \mathcal{P}(\{\lambda_i\}_{i=1}^{|\cX|})}\ \sum_{i=1}^{|\mathcal{T}|} \frac{\lambda^\prime_i}{n_0+\sum_{j\in t_i} n_j}+\sum_{i=|\mathcal{T}+1|}^{|\cX|}\frac{\lambda^\prime_i}{n_0},
\end{align}
where $\mathcal{P}(\{\lambda_i\}_{i=1}^{|\cX|})$ denotes the set of all possible permutations of $\{\lambda_1,\cdots,\lambda_{|\cX|}\}$. The solution of problem \eqref{eq:asd} is given in the following theorem. Without loss of generality, we rank the elements of $\mathcal{T}$ such that $\sum_{i \in t_1}n_i\ge \sum_{i \in t_2}n_i\ge \cdots\ge \sum_{i \in t_{|\mathcal{T}|}}n_i $.
\begin{theorem}
	\label{thm:wsd}
	Let $\{\lambda^*_i\}$ be the arguments that minimizing the objective of problem \eqref{eq:asd}, and then $\lambda^*_i=\lambda_i$.
\end{theorem}
\begin{algorithm}[t]
	\caption{Partition Searching Algorithm}
	\begin{algorithmic}[1]
		\STATE {\bfseries Input:} $\{\lambda_i\}_{i=1}^{|\cX|}$, $\{\bv_i\}_{i=1}^{|\cX|}$, and $\{n_i\}_{i=1}^{k}$\\
		\STATE $R\gets \sum_{i=1}^{|\cX|}\frac{\lambda_i}{n_0}$, $\mathcal{T}_0\gets \emptyset$\\
		\STATE {\bfseries for} $\mathcal{T}\in\mathcal{Q}$ {\bfseries do}\\
		\STATE \quad {Sort} $\mathcal{T}$ s.t. $\sum_{i \in t_1}n_i\ge \sum_{i \in t_2}n_i\ge \cdots\ge \sum_{i \in t_{|\mathcal{T}|}}n_i$\\
		\STATE \quad $R^\prime\gets \sum_{i=1}^{|\mathcal{T}|} \frac{\lambda_i}{n_0+\sum_{j\in t_i} n_j}+\sum_{i=|\mathcal{T}|+1}^{|\cX|}\frac{\lambda_i}{n_0}$
		\STATE \quad {\bfseries if} $R^\prime < R$ {\bfseries then}
		\STATE \quad \quad $R\gets R^\prime$, $\mathcal{T}_0\gets \mathcal{T}$
		\STATE {\bfseries end}\\
		\STATE {\bfseries for} $t_i\in \mathcal{T}_0$ {\bfseries do}
		\STATE \quad {\bfseries for} $a\in t_i$ {\bfseries do}
		\STATE \quad\quad $\bmu_{a}\gets \bv_i$
		\STATE \quad {\bfseries end}\\
		\STATE {\bfseries end}
		\STATE \bfseries{return} $\{\bmu_i\}_{i=1}^k$
	\end{algorithmic}
	\label{alg:1}
\end{algorithm}
	
With \thmref{thm:wsd}, the solution of problem \eqref{eq:oiu} lies in comparing the minimal EPRs for all possible partitions. Let $\mathcal{Q}$ be the collection of all possible partitions of $\{1,2,\cdots,k\}$. Such result can be summarized as Algorithm~\ref{alg:1}, whose outputs are the statistic functions as desired in problem \eqref{eq:nhg}. Moreover, the complexity of \algoref{alg:1} is the number of possible partitions of  $\{1,2,\cdots,k\}$, which is called Bell number~\cite{bel}, denoted as $B_k$. It has been found that $B_k\sim O((\frac{k}{\log k})^k)$~\cite{bell}, which can be smaller than the complexity $O(k^k)$ of trivial methods. 
\section{Vector transmission}
\label{sec:4}
Similar to the procedures in \secref{sec:3}, we first provide the maximum likelihood estimator to solve step (i) of problem \eqref{eq:que} as follows.
\begin{align}
	\label{eq:plk}
	\tilde{\bphi}_{\mathtt{ML}}=&\left(n_0\mathbf{I}+\sum_{i=1}^k n_i\bm{F}_i(\bm{F}_i^\T\bm{F}_i)^{-1}\bm{F}_i^\T\right)^{-1}\notag\\ &\quad \left(n_0\hat{\bphi}_0+\sum_{i=1}^k\bm{F}_i(\bm{F}_i^\T\bm{F}_i)^{-1}\bm{F}_i^\T\hat{\bphi}_i\right).
\end{align}
The matrix $\bm{F}_i^\T\bm{F}_i$ is not singular here, and otherwise the statistic $\hat{\bphi}_i\bm{F}_i$ could be equivalent to a lower-dimensional one. We without loss of generality assume that $\bm{F}_1^\T \bm{F}_1=\bm{F}_k^\T \bm{F}_k=\mathbf{I}$, which comes from the fact that we can do the linear transformation $\bm{F}_i(\bm{F}_i^\T\bm{F}_i)^{-\frac{1}{2}}$ for arbitrary $\bm{F}_i$. Let $\bm{F}_i=[\bmu_i^{(1)},\cdots,\bmu_i^{(m)}]$, and then step (ii) of problem \eqref{eq:que} becomes
\begin{align}
	\label{eq:qxcg}
	\min_{\{\{\bmu^{(j)}_i\}_{j=1}^m\}_{i=1}^k}&\ \tr\left[\mathbf{H}\left(n_0\mathbf{I}+\sum_{i=1}^k\sum_{j=1}^m n_i\bmu^{(j)}_i\bmu_i^{(j)\T}\right)^{-1}\right]\notag\\ 
	\mathrm{s.t.}&\ \bmu^{(j)\T}_i\bmu^{(j^\prime)}_i=\mathds{1}\{j=j^\prime\}
\end{align}
Similar to \thmref{thm:2}, we have the following characterization of the results in problem \eqref{eq:qxcg}.
\begin{corollary}
\label{col:1}
	Let $\{\{\bmu^{(j)*}_i\}_{j=1}^m\}_{i=1}^k$ be the optimal solution of \eqref{eq:qxcg}, and then
	\begin{align}
		\bmu_i^{(j)*}\in \{\bv_1,\bv_2, \cdots,\bv_{|\cX|}\},\ \forall i,j.
	\end{align}
	\end{corollary}

Corollary~\ref{col:1} implies that problem \eqref{eq:qxcg} is still to allocate different directions of eigenvectors to the entries $\bmu_i^{(j)}$ of different statistic functions. Additionally, we can still find the optimal statistic functions according to an algorithm similar to \algoref{alg:1}. The only difference lies in that for the case of scalar transmission, we consider the partition of the index set $\{1,2,\cdots,k\}$, where each index $i$ could appear once. For the case of vector transmission, we request each index appears $m$ times, where the $m$-th partition is defined as follows.
\begin{definition}
	A $m$-th partition $\mathcal{T}_m$ of a set $T$ satisfies (1) $\emptyset\notin \mathcal{T}_m$, (2) $\cup_{A\in \mathcal{T}_m}=T$, and (3) for all $t\in T$, $\sum_{A\in \mathcal{T}_m}\mathds{1}\{t\in A\}=m$.
\end{definition}

\begin{algorithm}[t]
	\caption{$m$-th Partition Searching Algorithm}
	\begin{algorithmic}[1]
		\STATE {\bfseries Input:} $\{\lambda_i\}_{i=1}^{|\cX|}$, $\{\bv_i\}_{i=1}^{|\cX|}$, and $\{n_i\}_{i=1}^{k}$\\
		\STATE $R\gets \sum_{i=1}^{|\cX|}\frac{\lambda_i}{n_0}$, $\mathcal{T}_0\gets \emptyset$, $\bm{F}_i\gets \emptyset$\\
		\STATE {\bfseries for} $\mathcal{T}_m=\{t_1,\cdots,t_{|\mathcal{T}_m|}\}\in\mathcal{Q}_m$ {\bfseries do}\\
		\STATE \quad {Sort} $\mathcal{T}_m$ s.t. $\sum_{i \in t_1}n_i\ge \sum_{i \in t_2}n_i\ge \cdots\ge \sum_{i \in t_{|\mathcal{T}_m|}}n_i$\\
		\STATE \quad $R^\prime\gets \sum_{i=1}^{|\mathcal{T}_m|} \frac{\lambda_i}{n_0+\sum_{j\in t_i} n_j}+\sum_{i=|\mathcal{T}_m|+1}^{|\cX|}\frac{\lambda_i}{n_0}$
		\STATE \quad {\bfseries if} $R^\prime < R$ {\bfseries then}
		\STATE \quad \quad $R\gets R^\prime$, $\mathcal{T}_0\gets \mathcal{T}$
		\STATE {\bfseries end}\\
		\STATE {\bfseries for} $t_i\in \mathcal{T}_0$ {\bfseries do}
		\STATE \quad {\bfseries for} $a\in t_i$ {\bfseries do}
		\STATE \quad\quad $\bm{F}_{a}\gets \bm{F}_{a}\cup\{\bv_i\}$
		\STATE \quad {\bfseries end}\\
		\STATE {\bfseries end}
		\STATE \bfseries{return} $\{\bm{F}_i\}_{i=1}^k$
	\end{algorithmic}
	\label{alg:2}
\end{algorithm}

Note that the standard partition in \secref{sec:3} can be viewed as the $1$-th partition of  set $\{1,2,\cdots,k\}$. With all these results, problem \eqref{eq:que} can be solved by finding the optimal $m$-th partition of the index set $\{1,\cdots,k\}$ ($m\ge k$). The procedures can be summarized in \algoref{alg:2}, where $\mathcal{Q}_m$ be the collection of all possible $m$-th partitions of $\{1,2,\cdots,k\}$. The outputs $\{\bm{F}_i\}_{i=1}^k$ are the collections of required statistic function entries from $k$ nodes, whose arrangement in row can be the solution of problem \eqref{eq:que}. Finally, the corresponding estimator for information vector $\bphi^*$ after knowledge transmission is as defined in \eqref{eq:plk}.
\section{Conclusion}
This paper studies the information transmission problem in distributed learning, where the design of the transmitted statistics is related to a singular vector decomposition problem. Under the asymptotic regime, the desired method allocates eigenvectors of the EPR norm matrix $\mathbf{H}$ to different statistic functions in consideration of the sample sizes and the eigenvalues. Note that this paper provides a general operation approach, and designing corresponding concrete algorithms for model training could be an interesting future direction.

\section*{Appendix}
\subsection{Proof of \propref{lem:2}}
\label{app:2}
We first provide some notations. We define the vector of loss function $l(x;\btheta)$ as
\begin{align}
    \bm{l}(\btheta)=\left[\sqrt{P_X(1)}l(1,\btheta),\cdots,\sqrt{P_X(|\mathcal{X}|)}l(|\mathcal{X}|,\btheta)\right]^T.
\end{align}
Then, the training loss as defined in \eqref{eq:ffdd} can be written as
\begin{align}
    \btheta^*=\argmin_{\btheta} \bphi^{*\T}\bm{l}(\btheta),
\end{align}
where we have
\begin{align}
    \bphi^{*\T}\nabla_{\btheta}\bm{l}(\btheta^*)=\bm{0}.
\end{align}
Simialrly,
\begin{align}
    \bphi^{\T}\nabla_{\btheta}\bm{l}(\hat{\btheta})=\bm{0}.
\end{align}

Note that
\begin{align}
    \hat{\bphi}^\T\nabla_{\btheta}\bm{l}(\hat{\btheta})=&(\bphi^*+\hat{\bphi}-\bphi^*)^\T\notag \\
    &\quad\left(\nabla_{\btheta}\bm l(\btheta^*)+\nabla^2_{\btheta}\bm l(\btheta^*)(\hat{\btheta}-\btheta^*)+\cdots\right).
\end{align}

It leads to

\begin{align}
(\hat{\btheta}-\btheta^*)=-\left[\bphi^{*\T}\nabla_{\btheta}^2 \bm l(\btheta^*)\right]^{\dagger}(\hat{\bphi}-\bphi^*)^\T\nabla_{\btheta} \bm l(\btheta^*)+o(\epsilon).
\end{align}

We can also get the Taylor series of the loss function with $\bphi$

\begin{align}
\bm l(\hat{\btheta})=&\bm l(\theta^*)+\nabla_{\btheta} \bm l(\btheta^*)^\T(\hat{\btheta}-\btheta^*)\notag \\
&\quad+\frac{1}{2}(\hat{\btheta}-\btheta^*)^\T\nabla_{\btheta}^2 \bm l(\btheta^*)(\hat{\btheta}-\btheta^*)+\cdots.
\end{align}

Finally, we can compute the testing loss \eqref{eq:kj} 
\begin{align}
	R(\hat{\btheta})&=\frac{1}{2}\E_{\mathtt{G}}\left[(\hat{\btheta}-\btheta^*)^\T[\bphi^{*\T}\nabla_{\btheta}^2 \bm l(\btheta^*)](\hat{\btheta}-\btheta^*)\right]+o(\epsilon^2)\notag\\
	&= \frac{1}{2}\E_{\mathtt{G}}\Big[(\hat{\bphi}-\bphi^*)^\T\nabla_{\btheta} \bm l(\btheta^*)\left[\bphi^{*\T}\nabla_{\btheta}^2 \bm l(\btheta^*)\right]^{\dagger}\notag\\
	&\quad\quad\qquad\qquad\nabla_{\btheta} \bm l(\btheta^*) ^\T(\hat{\bphi}-\bphi^*)\Big]+o(\epsilon^2),
\end{align}
where \propref{lem:2} is proved.
\subsection{Proof of \thmref{thm:1}}
	\label{pf:1}
The MLE $\tilde{\bphi}_{\mathtt{ML}}$ corresponds to the mean square error 
\begin{align}
\label{eq:mnb}
    &\mathbb{E}_{\mathtt{G}}\left[(\tilde{\bphi}_{\mathtt{ML}}-\bphi^*)^\T\mathbf{H}(\tilde{\bphi}_{\mathtt{ML}}-\bphi^*)\right]\notag\\
    &=\tr\left[\mathbf{H}\left(n_0\mathbf{I}+\sum_{i=1}^kn_i\bmu_i\bmu_i^\T\right)^{-1}\right]
\end{align}
While for all possible estimator $\tilde{\bphi}$, the Cramer-Rao bound for its mean square error hold:
\begin{align}
\label{eq:cxd}
    &\mathbb{E}_{\mathtt{G}}\left[(\tilde{\bphi}-\bphi^*)^\T\mathbf{H}(\tilde{\bphi}-\bphi^*)\right]\notag\\
    &\ge\tr\Bigg\{\mathbb{E}\Bigg[\nabla^2_{ H^\frac{1}{2}\bphi^*}\Big(\log\mathbb{P}(\hat{\bphi}_0;\bphi^*)\notag\\
    &\quad\quad\quad\quad\quad\quad\quad\quad+\sum_{i=1}^k\log\mathbb{P}(\hat{\bphi}_i^T\bmu_i;\bphi^*)\Big)\Bigg]^{-1}\Bigg\}\notag\\
    &=\tr\left[\mathbf{H}\left(n_0\mathbf{I}+\sum_{i=1}^kn_i\bmu_i\bmu_i^\T\right)^{-1}\right]
\end{align}

Note that error \eqref{eq:mnb} takes the Cramer-Rao bound \eqref{eq:cxd} and thus \thmref{thm:1} is proved.
\subsection{Proof of \thmref{thm:2}}
	\label{pf:2}
First, we consider he method of Lagrange multipliers and derive the following equations. For all $i\in\{1,2,\cdots,k\}$
\begin{align}
\label{eq:gtrf}
    \left(n_0\mathbf{I}+\sum_{j=1}^kn_j\bmu_j\bmu_j^\T\right)^{-1}\!\!\!\!\!\mathbf{H}\left(n_0\mathbf{I}+\sum_{j=1}^kn_j\bmu_j\bmu_j^\T\right)^{-1}\!\!\!\!\!\bmu_i=\xi_i\bmu_i,  
\end{align}
where $\{\xi_i\}_{i=1}^k$ are the multipliers.
Apparently, $\bmu_i\in \{\bv_1,\bv_2, \cdots,\bv_{|\cX|}\},\ \forall i\in\{1,2,\cdots,k\}$ satisfies this equation.
Note that equation \eqref{eq:gtrf} is equivalent to
\begin{align}
\label{eq:ploi}
    &\left(n_0\mathbf{I}+\sum_{j=1}^kn_j\bmu_j\bmu_j^\T\right)\!\!\mathbf{H}^{-1}\!\!\left(n_0\mathbf{I}+\sum_{j=1}^kn_j\bmu_j\bmu_j^\T\right)\!\bmu_i=\!\frac{1}{\xi_i}\bmu_i\notag\\
    &\Rightarrow\mathbf{H}^{-\frac 12}\left(n_0\mathbf{I}+\sum_{j=1}^kn_j\bmu_j\bmu_j^\T\right)\bmu_i=\frac{1}{\sqrt{\xi_i}}\bmu_i
\end{align}
It implies that vectors $\{\bmu_i\}_{i=1}^k$ are the eigenvectors of $\left(n_0\mathbf{I}+\sum_{j=1}^kn_j\bmu_j\bmu_j^\T\right)\mathbf{H}^{-1}\left(n_0\mathbf{I}+\sum_{j=1}^kn_j\bmu_j\bmu_j^\T\right)$.

Note that without loss of generality, $n_0, n_1,\cdots, n_k$ can be chosen such that the eigenvalues of this matrix are different from each other. As the eigenvectors of the same matrix, $\{\bmu_i\}_{i=1}^k$ can be parallel or perpendicular to each other. Let $\mathcal{A}_i$ be the index set that $\forall a\in \mathcal{A}_i$, $\bmu_a=\bmu_i$. Then, $\forall a\notin \mathcal{A}_i$, $\bmu_a\bot\bmu_i$. With these properties we have 
\begin{align}
    \mathbf{H}^{-\frac 12}\bmu_i\parallel\bmu_i,
\end{align}
which means $\bmu_i$ is the eigenvector of matrix $\mathbf{H}^{-\frac12}$.
Note that the eigenvectors of $\mathbf{H}$ and $\mathbf{H}^{-\frac12}$ are the same, and thus \thmref{thm:2} is proved.
\bibliographystyle{IEEEtran}
\bibliography{IEEEabrv,ref}

\begin{thebibliography}{10}
\providecommand{\url}[1]{#1}
\csname url@samestyle\endcsname
\providecommand{\newblock}{\relax}
\providecommand{\bibinfo}[2]{#2}
\providecommand{\BIBentrySTDinterwordspacing}{\spaceskip=0pt\relax}
\providecommand{\BIBentryALTinterwordstretchfactor}{4}
\providecommand{\BIBentryALTinterwordspacing}{\spaceskip=\fontdimen2\font plus
\BIBentryALTinterwordstretchfactor\fontdimen3\font minus
  \fontdimen4\font\relax}
\providecommand{\BIBforeignlanguage}[2]{{%
\expandafter\ifx\csname l@#1\endcsname\relax
\typeout{** WARNING: IEEEtran.bst: No hyphenation pattern has been}%
\typeout{** loaded for the language `#1'. Using the pattern for}%
\typeout{** the default language instead.}%
\else
\language=\csname l@#1\endcsname
\fi
#2}}
\providecommand{\BIBdecl}{\relax}
\BIBdecl

\bibitem{VerbraekenWKKVR20}
J.~Verbraeken, M.~Wolting, J.~Katzy, J.~Kloppenburg, T.~Verbelen, and J.~S.
  Rellermeyer, ``A survey on distributed machine learning,'' \emph{{ACM}
  Comput. Surv.}, vol.~53, no.~2, pp. 30:1--30:33, 2020.

\bibitem{li2020distributed}
H.~Li, Q.~L{\"u}, Z.~Wang, X.~Liao, and T.~Huang, \emph{Distributed
  Optimization: Advances in Theories, Methods, and Applications}.\hskip 1em
  plus 0.5em minus 0.4em\relax Springer, 2020.

\bibitem{LiuHZLJXD22}
J.~Liu, J.~Huang, Y.~Zhou, X.~Li, S.~Ji, H.~Xiong, and D.~Dou, ``From
  distributed machine learning to federated learning: a survey,'' \emph{Knowl.
  Inf. Syst.}, vol.~64, no.~4, pp. 885--917, 2022.

\bibitem{ImteajTWLA22}
A.~Imteaj, U.~Thakker, S.~Wang, J.~Li, and M.~H. Amini, ``A survey on federated
  learning for resource-constrained iot devices,'' \emph{{IEEE} Internet Things
  J.}, vol.~9, no.~1, pp. 1--24, 2022.

\bibitem{Bottou10}
L.~Bottou, ``Large-scale machine learning with stochastic gradient descent,''
  in \emph{{COMPSTAT}}, 2010.

\bibitem{h17qsgd}
D.~Alistarh, D.~Grubic, J.~Li, R.~Tomioka, and M.~Vojnovic, ``{QSGD:}
  communication-efficient {SGD} via gradient quantization and encoding,'' in
  \emph{Advances in Neural Information Processing Systems (NIPS)}, 2017.

\bibitem{Wangni18sparse}
J.~Wangni, J.~Wang, J.~Liu, and T.~Zhang, ``Gradient sparsification for
  communication-efficient distributed optimization,'' in \emph{NIPS}, 2018.

\bibitem{JiangA18}
P.~Jiang and G.~Agrawal, ``A linear speedup analysis of distributed deep
  learning with sparse and quantized communication,'' in \emph{NeurIPS}, 2018.

\bibitem{BasuDKD20}
D.~Basu, D.~Data, C.~Karakus, and S.~N. Diggavi, ``Qsparse-local-sgd:
  Distributed {SGD} with quantization, sparsification, and local
  computations,'' \emph{{IEEE} J. Sel. Areas Inf. Theory}, 2020.

\bibitem{Stich19}
S.~U. Stich, ``Local {SGD} converges fast and communicates little,'' in
  \emph{International Conference on Learning Representations, {ICLR}}, 2019.

\bibitem{SpiridonoffOP21}
A.~Spiridonoff, A.~Olshevsky, and Y.~Paschalidis, ``Communication-efficient
  {SGD:} from local {SGD} to one-shot averaging,'' in \emph{NeurIPS}, 2021.

\bibitem{GorbunovHR21}
E.~Gorbunov, F.~Hanzely, and P.~Richt{\'{a}}rik, ``Local {SGD:} unified theory
  and new efficient methods,'' in \emph{International Conference on Artificial
  Intelligence and Statistics,{AISTATS}}, 2021.

\bibitem{dis}
M.~F. Balcan, A.~Blum, S.~Fine, and Y.~Mansour, ``Distributed learning,
  communication complexity and privacy,'' in \emph{Conference on Learning
  Theory}.\hskip 1em plus 0.5em minus 0.4em\relax JMLR Workshop and Conference
  Proceedings, 2012, pp. 26--1.

\bibitem{csi}
I.~Csisz{\'a}r, ``The method of types [information theory],'' \emph{IEEE
  Transactions on Information Theory}, vol.~44, no.~6, pp. 2505--2523, 1998.

\bibitem{indicator}
W.~Feller, \emph{An introduction to probability theory and its applications,
  vol 2}.\hskip 1em plus 0.5em minus 0.4em\relax John Wiley \& Sons, 2008.

\bibitem{nphard}
J.~Van~Leeuwen, \emph{Handbook of theoretical computer science (vol. A)
  algorithms and complexity}.\hskip 1em plus 0.5em minus 0.4em\relax Mit Press,
  1991.

\bibitem{bel}
H.~Becker and J.~Riordan, ``The arithmetic of bell and stirling numbers,''
  \emph{American journal of Mathematics}, vol.~70, no.~2, pp. 385--394, 1948.

\bibitem{bell}
D.~Berend and T.~Tassa, ``Improved bounds on bell numbers and on moments of
  sums of random variables,'' \emph{Probability and Mathematical Statistics},
  vol.~30, no.~2, pp. 185--205, 2010.

\end{thebibliography}


\end{document}